\documentclass[fleqn,usenatbib]{mnras}
\usepackage{newtxtext,newtxmath}
\usepackage[T1]{fontenc}
\DeclareRobustCommand{\VAN}[3]{#2}
\let\VANthebibliography\thebibliography
\def\thebibliography{\DeclareRobustCommand{\VAN}[3]{##3}\VANthebibliography}

\usepackage{graphicx}	% Including figure files
\usepackage{amsmath}	% Advanced maths commands

\newcommand{\orcid}[1]{\href{https://orcid.org/#1}{\includegraphics[scale=0.08]{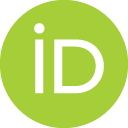}}}

\title[Chemical \& Thermal structure of Galaxy Groups]{Distribution of Metals and Multi-Temperature Gas in the Cores of Nearby Galaxy Groups}
\author[D. Chatzigiannakis et al.]{
Dimitris Chatzigiannakis\orcid{0009-0008-3247-9489},$^{1,2}$\thanks{E-mail: chatzigiannakis@mpia.de}
Aurora Simionescu\orcid{0000-0002-9714-3862},$^{3,4,5}$
Fran\c{c}ois Mernier\orcid{0000-0002-7031-4772}$^{6,7,8}$
\\
$^{1}$Max-Planck-Institut für Astronomie, Königstuhl 17, D-69117 Heidelberg, Germany \\
$^{2}$Heidelberg University, Astronomy and Physics department, Im Neuenheimer Feld 226, D-69120 Heidelberg, Germany\\
$^{3}$SRON Netherlands Institute for Space Research, Niels Bohrweg 4, 2300 RA Leiden, The Netherlands\\
$^{4}$Leiden Observatory, Leiden University, Niels Bohrweg 2, 2300 RA Leiden, The Netherlands\\
$^{5}$Kavli Institute for the Physics and Mathematics of the Universe (WPI), The University of Tokyo, Kashiwa, Chiba 277-8583, Japan\\
$^{6}$ NASA Goddard Space Flight Center, Code 662, Greenbelt, MD 20771, USA\\
$^{7}$ Department of Astronomy, University of Maryland, College Park, MD 20742-2421, USA\\
$^{8}$ IRAP, CNRS, Université de Toulouse, CNES, UT3-UPS, Toulouse, France
}

\date{Accepted XXX. Received YYY; in original form ZZZ}

\pubyear{2025}

\begin{document}
\label{firstpage}
\pagerange{\pageref{firstpage}--\pageref{lastpage}}
\maketitle

\begin{abstract}
Previous studies of galaxy clusters have focused extensively on the effects of active galactic nuclei (AGN) feedback on the chemical evolution of the intra-cluster medium (ICM). However, similar studies on the atmospheres of lower mass systems, such as galaxy groups and giant ellipticals, remain limited. In this work, we present a systematic analysis of the chemical and multi-temperature structure of the intra-group medium (IGrM), using a subsample of nearby galaxy groups and ellipticals from the CHEERS catalogue. By comparing areas with and without AGN feedback related features, such as cavities or extended radio lobes, we find clear evidence of an excess of multi-phase gas along the path of recent AGN feedback. However, its distribution exceeds the length of the radio lobes, since we recover a non-negligible amount of multi-phase gas at larger radii. In contrast to the clear asymmetry in the thermal structure, we find no directional enhancement in the distribution of Fe, with little to no differences in the Fe abundances of the on- and off-lobe directions. Our analysis suggests that the metals in the IGrM of our targets are well-mixed and decoupled from the effects of recent AGN feedback, as indicated by radio-lobes and cavities.
\end{abstract}


%%%%%%%%%%%%%%%%%%%%%%%%%%%%%%%%%%%%%%%%%%%%%%%%%%

%%%%%%%%%%%%%%%%% BODY OF PAPER %%%%%%%%%%%%%%%%%%

\section{Introduction}
Feedback from active galactic nuclei (AGN), referring to the interplay between the energy released by a super massive black hole (SMBH) and its surrounding gas, is an integral part of our understanding of galaxy evolution  \citep[e.g.][]{Silk98,Gebhardt00}. However, despite its ubiquitous use across galaxy evolution models \citep[e.g.][]{Schaye14,Vogelsberger14,Pillepich18,Dave19,Schaye23}, there are multiple, sometimes conflicting, implementations of it, with its exact nature remaining an open question. Therefore, studying the way AGN feedback interacts with its environment and placing observational constraints plays an important role in distinguishing between said implementations.

One observable that can be used to test feedback mechanisms is the large scale distribution of metals. Previous studies have demonstrated a direct link between AGN feedback and the observed chemical structure of the intra-cluster medium (ICM) \citep[e.g.][]{Werner13,Urban17,Biffi18,Mernier18,Gastaldello21}, with strong evidence in favour of an increased amount of metals along its direction, as indicated by the presence of radio jets/lobes and X-ray cavities \citep[e.g][]{Simionescu08,Simionescu09,Kirkpatrick11,Kirkpatrick15}. This highlights a mechanism by which heavier metals produced by the brightest cluster galaxy's (BCG) stellar population are uplifted by AGN feedback, contributing to the chemical enrichment of the ICM. 

However, the effects of AGN feedback are not as well understood when examining the hot atmospheres of less massive systems, such as galaxy groups and giant ellipticals. Studies of the chemistry of the intra-group medium (IGrM) reveal various underlying patterns, ranging from strong Fe enhancements along radio jets and lobes, similar to what we see in clusters, to no- or even anti-correlations with the direction of known features related to AGN feedback \citep[e.g.][]{O'Sullivan05,O'Sullivan11,Lagana15,Randall15,GendronM17,Su19}. This discrepancy could be a product of the Fe bias \citep{Buote99}, systematically leading to lower best-fit Fe abundances if a multi-phased medium, such as the gas along radio lobes \citep[e.g.][]{Simionescu08}, is modelled with a single temperature component. In \citet{Chatzigiannakis22} we examined this hypothesis by modelling the multi-phase gas of NGC~5813; while we have found an extended distribution of cooler gas along the direction of the radio lobes, the Fe abundance of the IGrM appeared to be rather uniform, with no evidence of an abundance enhancement. However, the limited number of such works on other lower mass systems, with various applied methodologies, did not allow us to draw any definite or more general conclusions.

In this pilot study we aim to study the effects of AGN feedback on the IGrM in a systematic way for the first time, by applying a consistent analysis method on a larger sample of galaxy groups and giant ellipticals. Primarily, we focus on the azimuthally resolved properties of the multi-phase thermal and chemical structure of the IGrM and how they correlate with the direction of AGN feedback. The structure of this paper proceeds as follows: In Section \ref{sec:data} we present our sample selection as well as the spectral modelling and fitting strategy used in this work. In Section \ref{sec:results} we highlight our main results, while our findings are discussed in greater detail in Section \ref{sec:discussion}. Finally, our main results are summarized in Section \ref{sec:conclusions}.

For the purposes of this analysis, we assume a standard $\Lambda$CDM cosmology ($\Omega_m=0.3$, $\Omega_\Lambda=0.7$ and $H_0=70~\rm{km~s^{-1}Mpc^{-1}}$). Uncertainties are given at 68\% confidence intervals (1$\sigma$), unless otherwise stated. Elemental abundances are estimated using the \citet{Asplund09} Solar abundance reference tables.

\section{Data Analysis} \label{sec:data}
\subsection{Observations and data preparation}
In this work, we use archival \textit{XMM-Newton} observations for a subset of 8 nearby cool-core galaxy groups and ellipticals from the CHEmical Enrichment Rgs Sample (CHEERS) catalogue. \citep{dePlaa17,Mernier16a}. We present a full list of our selected targets, their properties, and the relevant observations used in this work in Table \ref{tab:sample}. We select our targets based on their "group" sub-sample classification and clear evidence of ongoing AGN feedback, in the form of X-ray cavities, as indicated in \citet{Panagoulia14}. We note that all of our targets, with the exception of NGC 4325, have extended radio emission in the form of radio lobes \citep{Giacintucci11,Kolokythas20}, that coincide with the location of the X-ray cavities, further supporting the notion that AGN feedback is currently present in those systems.

For the reduction of the European Photon Imaging Camera (EPIC) MOS1, MOS2 and pn data of the relevant observations, we follow the process described in \citet{Mernier17}. Namely, we extract the event files from MOS and pn data using the XMM Science Analysis System (SAS) v14.0 \texttt{emproc} and \texttt{epproc} commands, respectively, and filter out each light-curve from soft-flare events, following \citet{Mernier17}. Similarly, we exclude point sources in our field of view, identified by \citet{Mernier17}, by masking a circular 10" region around their reported surface brightness peak, detected via the \texttt{edetect\_chain} and verified by eye.

For each observation, we extract the MOS~1, MOS~2 and pn spectra of regions defined by a radial and an azimuthal bin. For the radial bins, we define a set of annuli of fixed angular sizes (0'-0.5', 0.5'-1, 1'-2', 2'-3', 3'-4' and 4'-6'), following \citet{Mernier17}, with the centre placed on the peak of the X-ray emission of the EPIC surface brightness images. We choose to exclude annuli at higher angular separations from our analysis due to low contribution of the IGrM in the spectrum with respect to the background (see \ref{sec:background}).

\begin{figure*}
    \centering 
    \includegraphics[height=5.5cm]{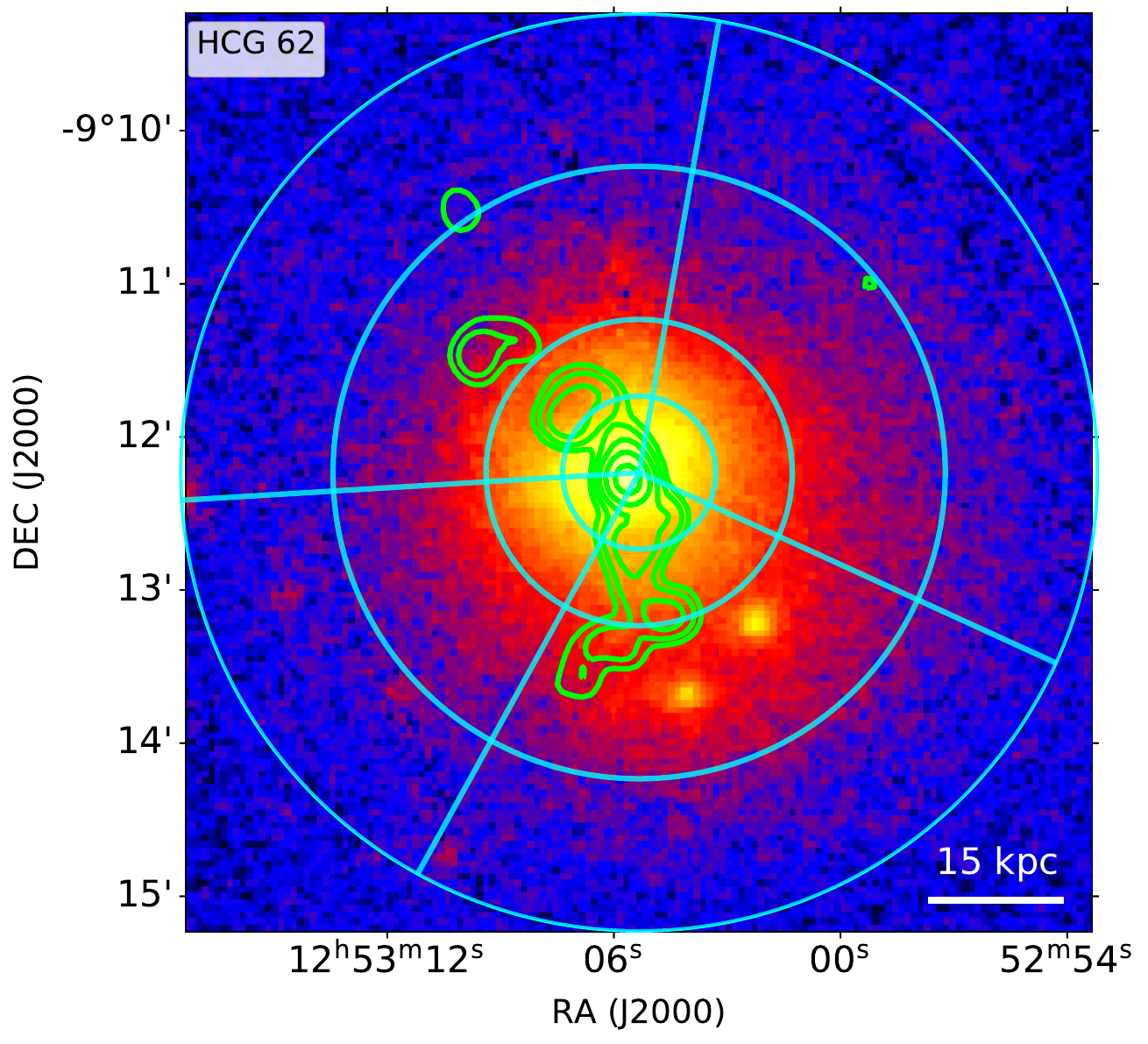}
    \includegraphics[height=5.5cm]{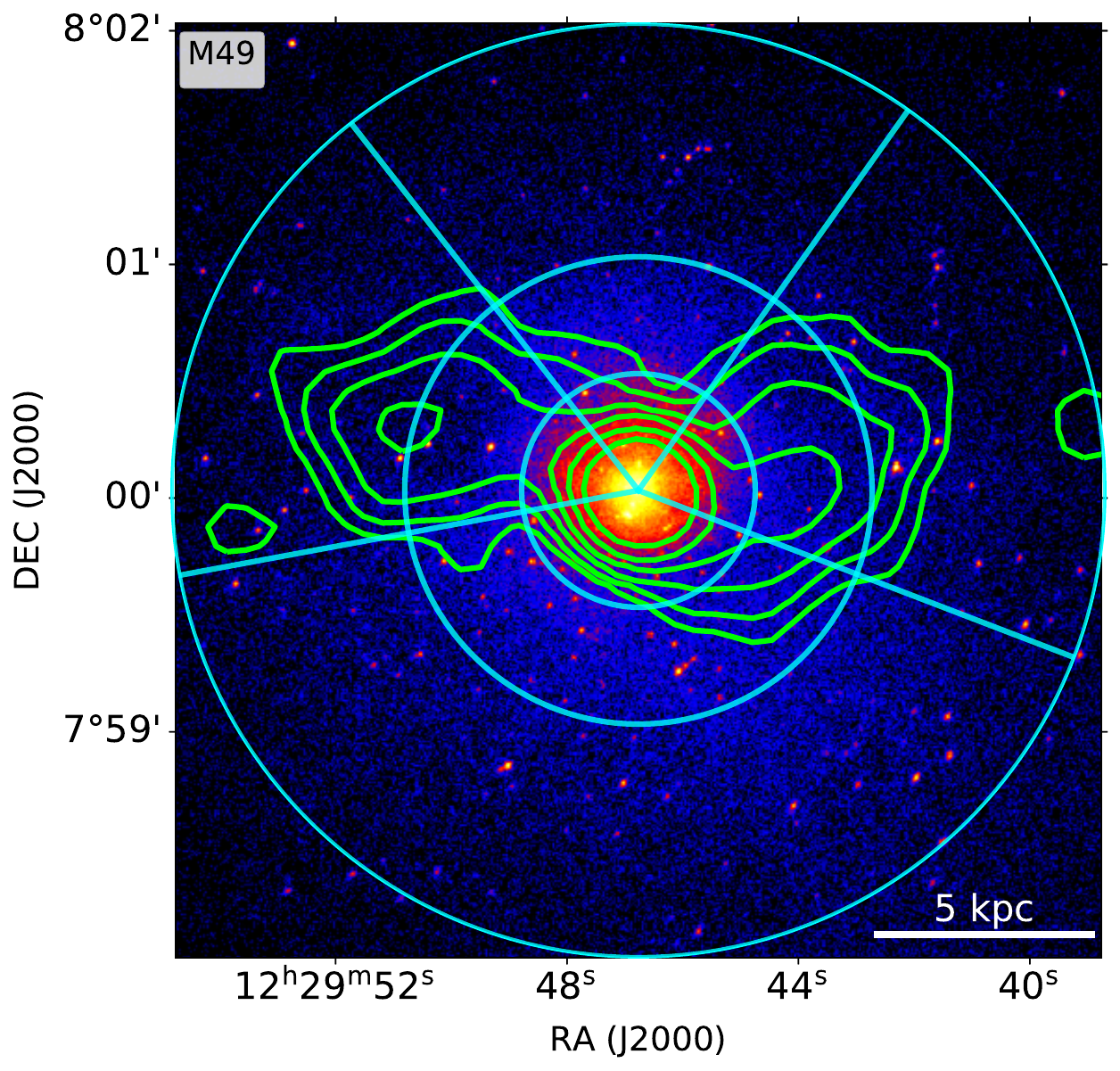}
    \includegraphics[height=5.5cm]{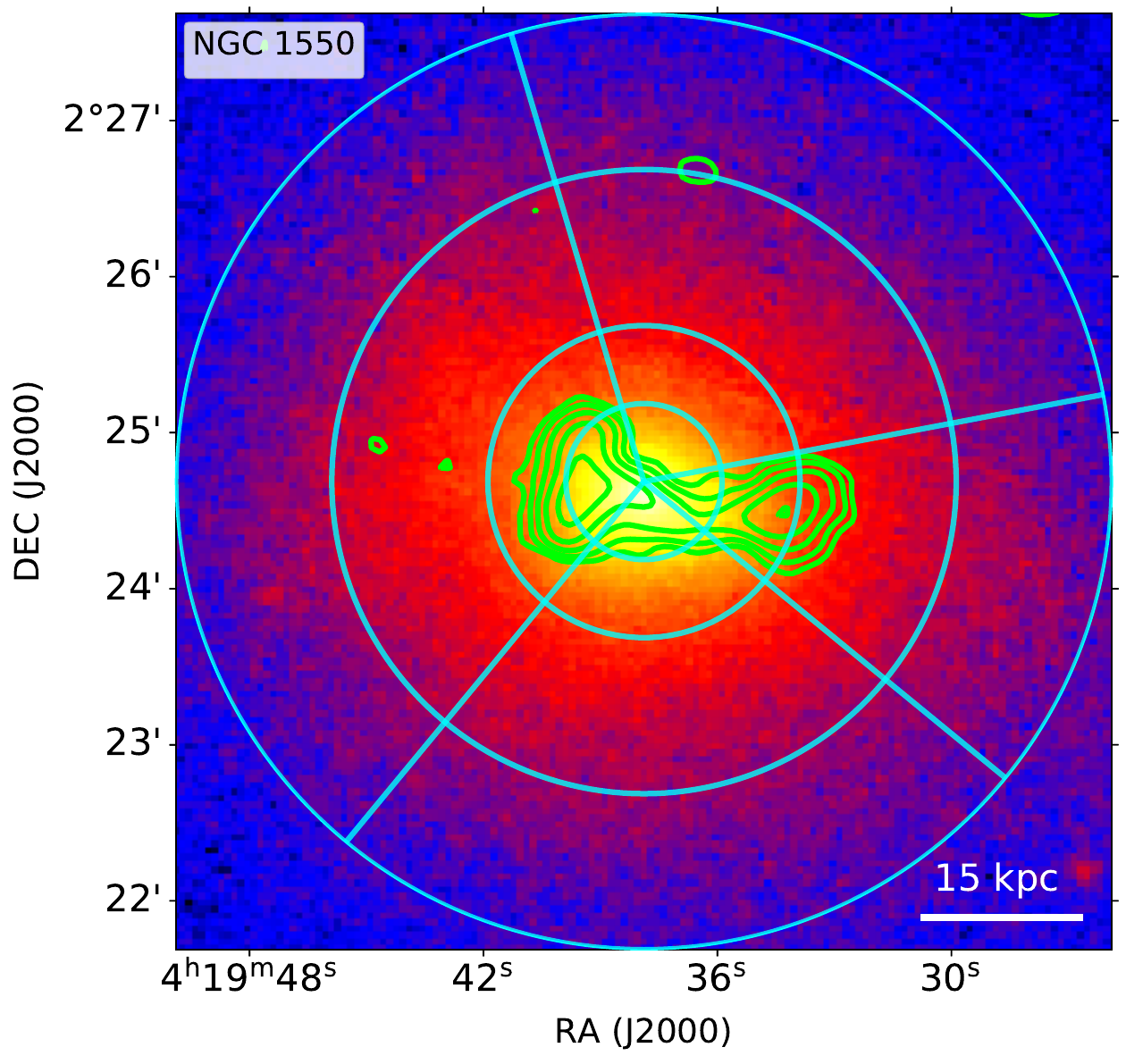}
    \medskip
    \includegraphics[height=5.48cm]{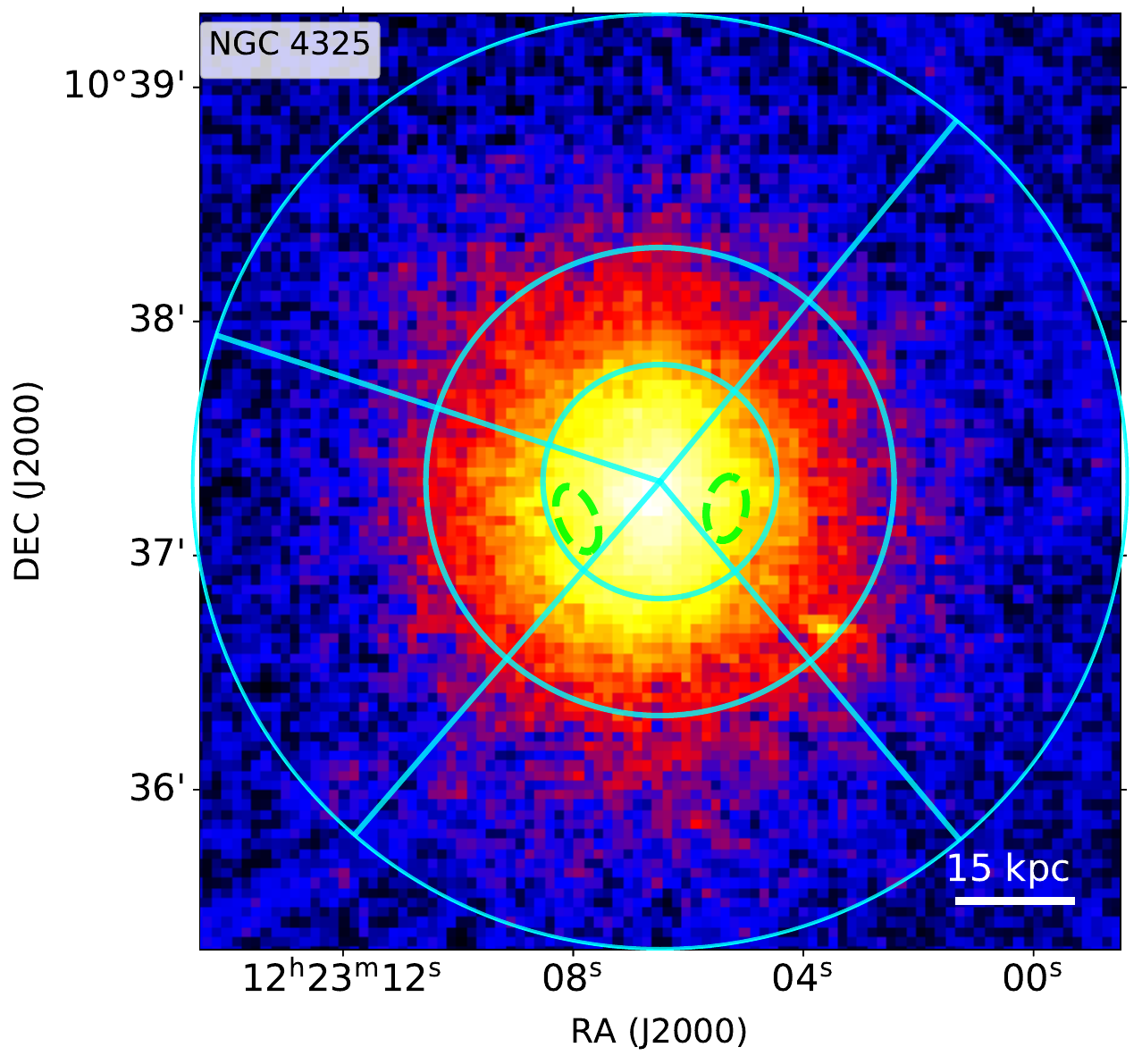}
    \includegraphics[height=5.48cm]{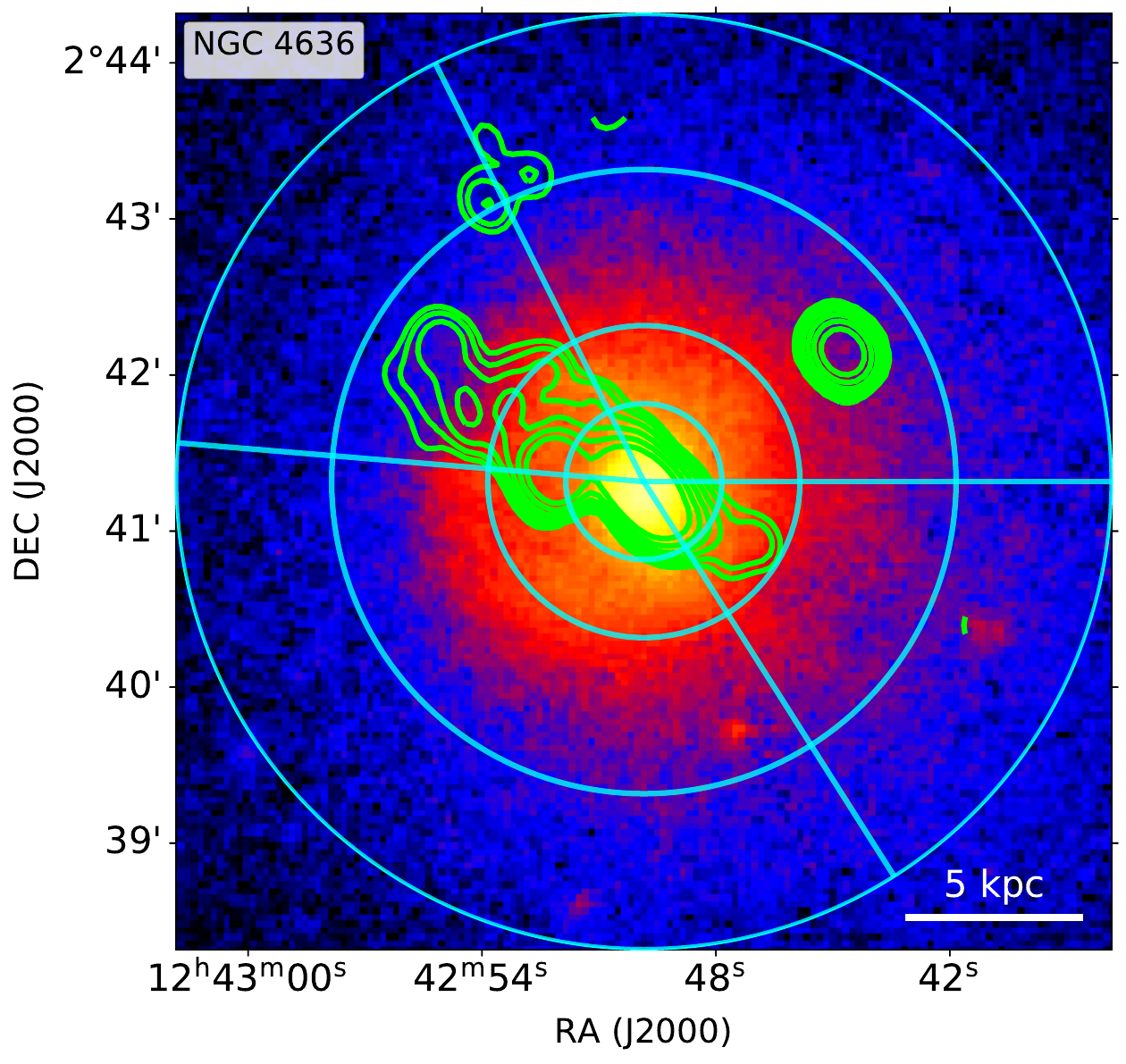}
    \includegraphics[height=5.48cm]{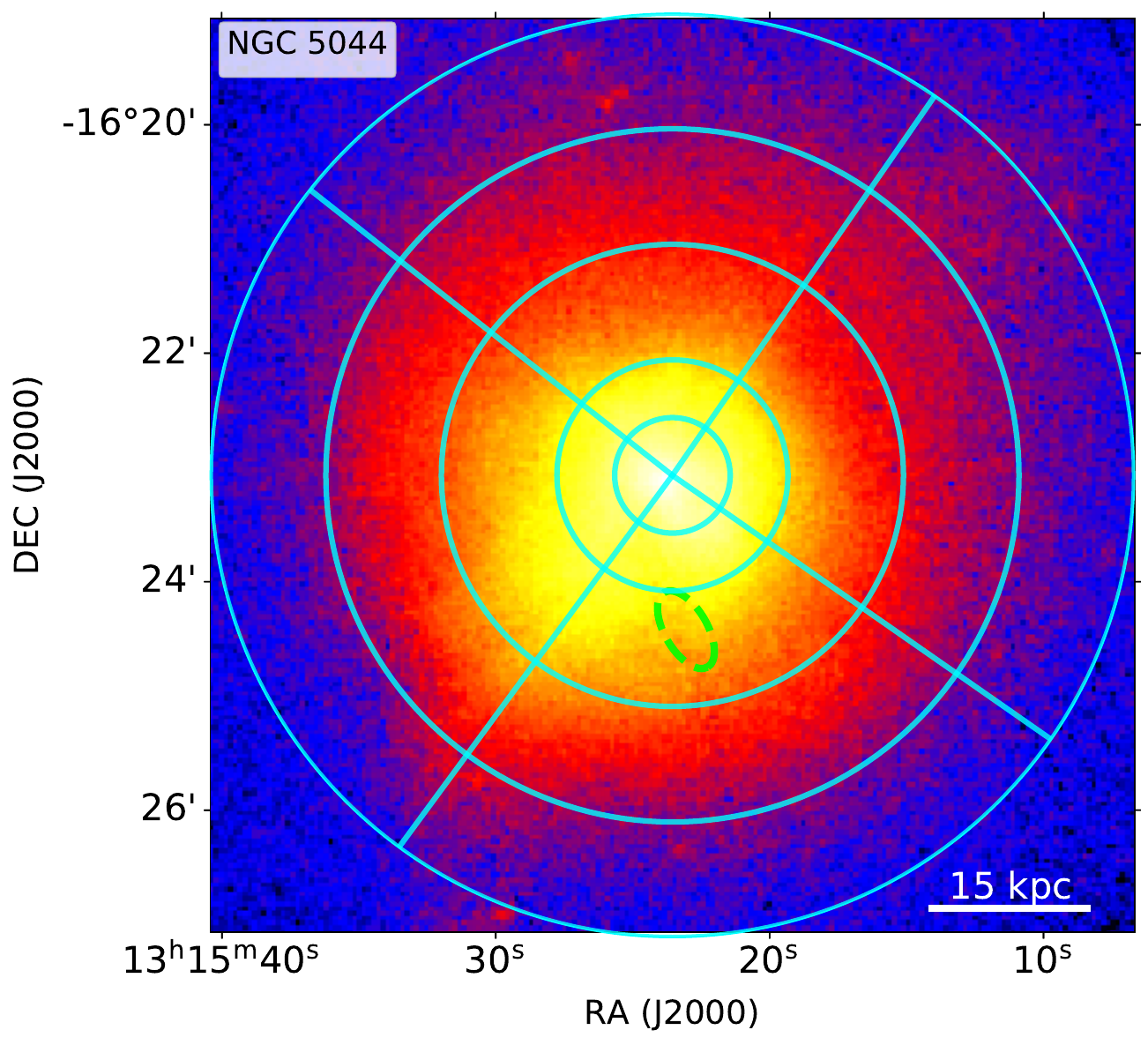}
    \medskip
    \includegraphics[height=5.5cm]{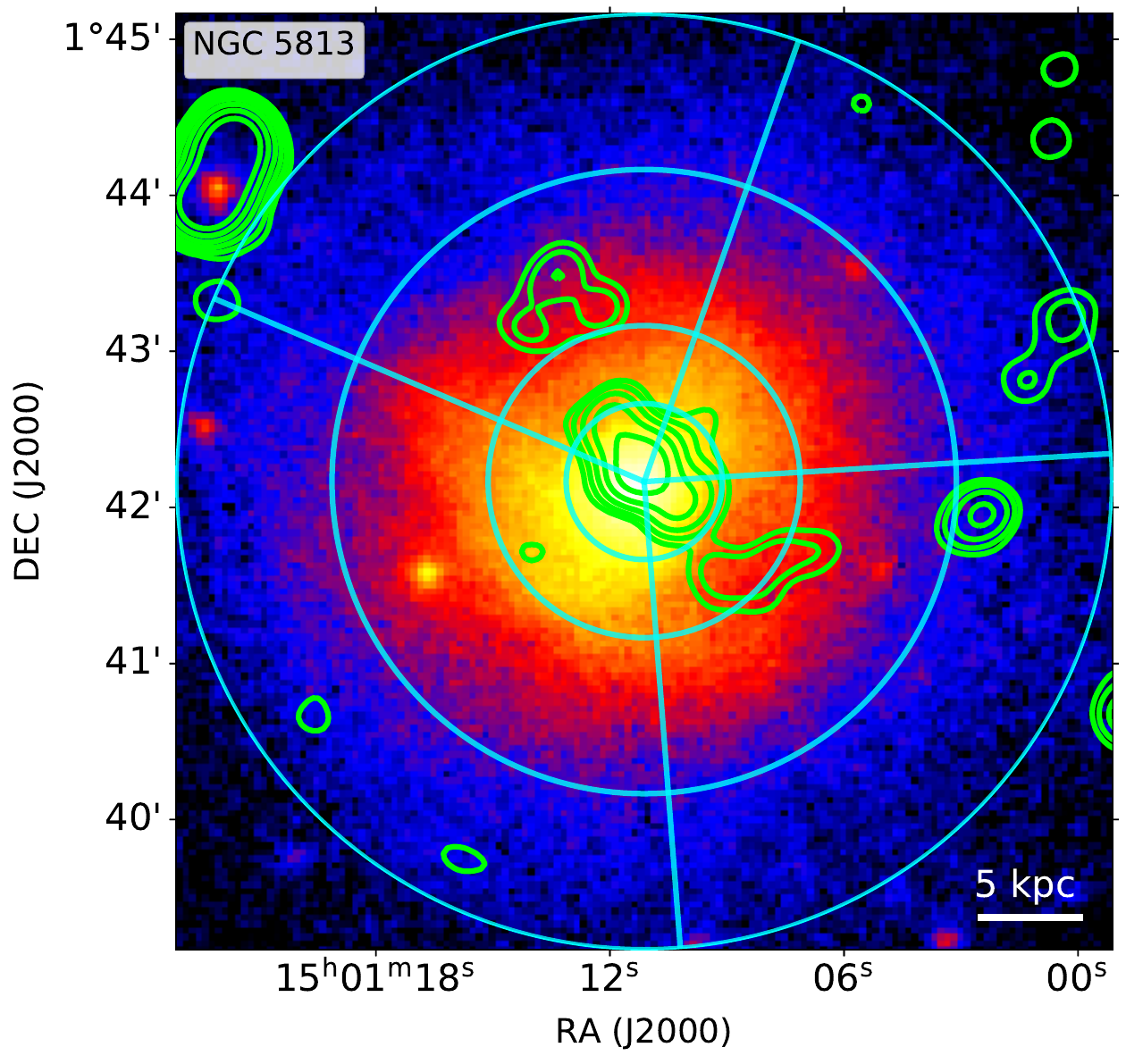}
    \includegraphics[height=5.5cm]{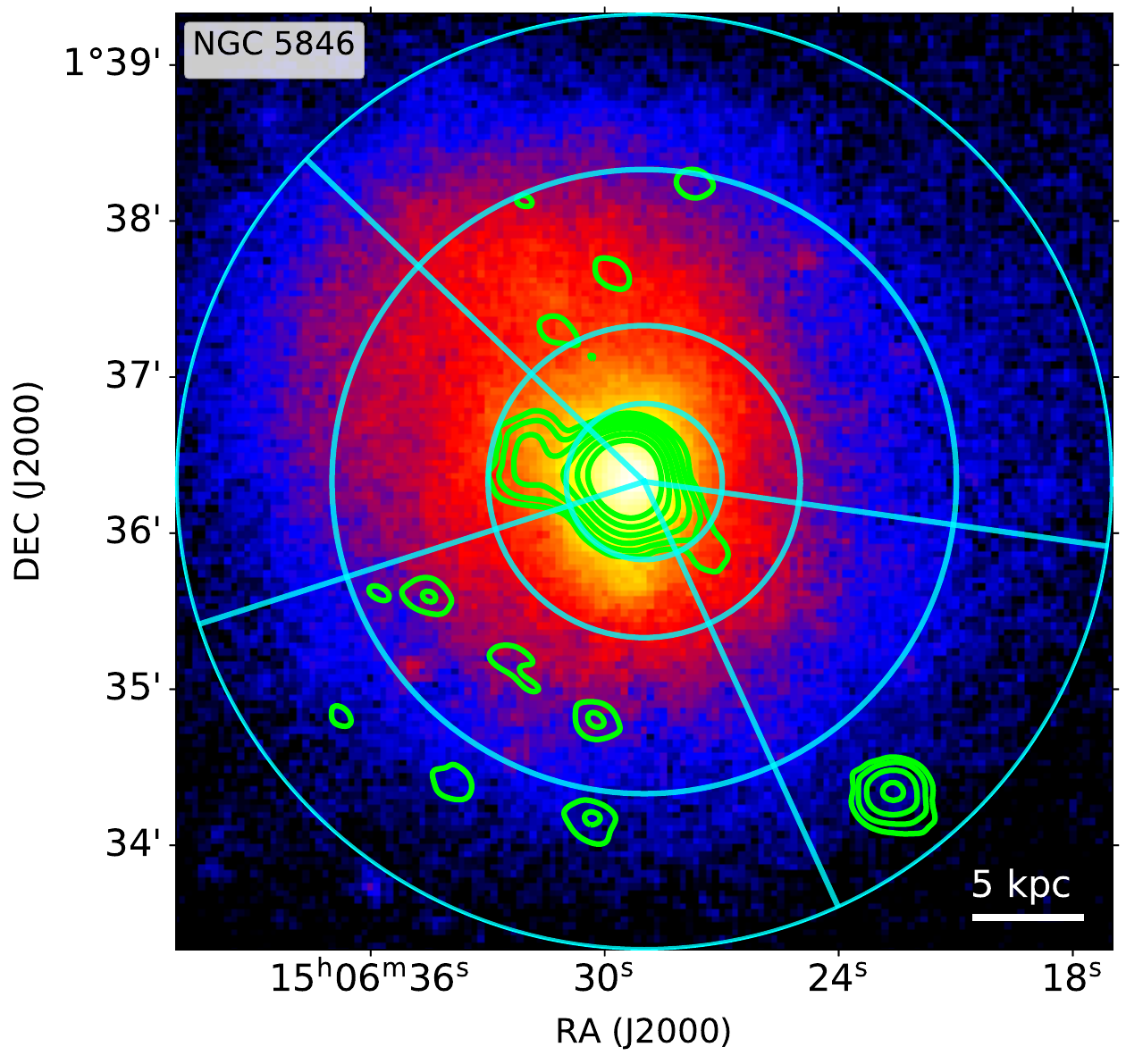}
\caption{Exposure-corrected, background subtracted, 0.3-2 keV combined EPIC MOS1, 2 and pn images, smoothed with a $\sigma=1.25"$ Gaussian kernel, for the CHEERS sample targets used in this work: HCG 62, M49/NGC~4472, NGC~1550, NGC~4325, NGC~4636, NGC~5044, NGC~5813 and NGC~5846 from left to right and from up to down. Green contours indicate \textit{GMRT} 236~MHz emission (\textit{GMRT} 610~MHz for NGC~5846) adopted from \citet{Giacintucci11} and \citet{Kolokythas20} for NGC~1550. If radio data were not available, the location of known X-ray cavities from \citet{Panagoulia14} has been indicated instead with dashed green ellipses. Cyan lines define the individual radial and azimuthal bins delimiting the on- and off-lobe directions.}
\label{fig:sample}
\end{figure*}

For the azimuthal bins, we follow the direction of each target's radio lobes and the location of known X-ray cavities, as shown in \autoref{fig:sample}. This way, we define a set of directions with and without direct evidence of ongoing AGN feedback, hereafter the on- and off-lobe directions, respectively. We note that, due to the complex morphology of said lobes, the on-lobe directions are not necessarily co-linear and, due to the differences in the opening angles of the radio lobes, the area of each azimuthal bin varies. We also point out that, primarily due to the morphology of the radio emission in the centres of our galaxy groups and ellipticals, some off-lobe regions at small radii coincide with areas where a significant radio detection is still present. 

Both the redistribution matrix and ancillary response files (RMF and ARF respectively) for each extracted spectrum are produced via the SAS tasks \texttt{rmfgen} and \texttt{arfgen}.

\subsection{Spectral modelling and analysis}
For the purposes of our analysis, we use the X-ray spectral-fitting program \texttt{XSPEC} \citep{Arnaud96}, and the AtomDB 3.0.9 atomic database. As mentioned, we extract multiple spectra for each defined region of interest, with a MOS~1, MOS~2 and a pn spectrum per available observation, that we fit simultaneously. We chose to fit our spectra within the 0.6-3.0~keV energy range, as it encompasses the Fe-L complex, which is the peak of the X-ray emission in galaxy groups and ellipticals, while avoiding modelling uncertainties around the detector's O edge.

In the following sections we describe our modelling of the thermal emission and background for our spectra, followed by our fitting strategy.

\subsubsection{Thermal Emission}
We describe the emission of the IGrM with an absorbed \texttt{vlognorm}\footnote{https://github.com/jeremysanders/lognorm} model. The model assumes a normal distribution of single temperature \texttt{apec} models in the natural logarithm space, around a central temperature. It describes the extent of the underlying multi-temperature structure via the width of the distribution ($\sigma_{\text{kT}}$), and converges to a single temperature model for decreasing widths. Our choice is motivated by the fact that a log-normal thermal distribution has been shown to provide a more accurate representation of the multi-phase ICM/IGrM's chemical and thermal properties both observationally \citep[e.g.][]{Simionescu09,Mernier17} and in simulations \citep[e.g.][]{Khedekar13,Zhuravleva13,Zhang24,Chatzigiannakis25}. 

Regarding the chemical abundances, in this work we are primarily interested in Fe, as our main proxy of the chemical enrichment of the IGrM. The abundances of all other elements are coupled to Fe, with the following exceptions: He/H is kept at 1 Solar; Si is free to vary, in order to account for possible remaining issues in the modelling of the Si K$\alpha$ instrumental line; Mg is free to vary since the Mg K$\alpha$ line is blended with some of the weak Fe-L lines and could bias our inferences of the multi-temperature structure if left frozen.

The model assumes a redshifted emission being absorbed by a neutral foreground along our line of sight, that we describe with the addition of a \texttt{phabs} component. Both the redshift of our targets and the equivalent hydrogen column density along the line of sight remain frozen to the values listed in Table \ref{tab:sample}.

\begin{table*}
    \centering
    \begin{tabular}{l c c c c}
    \hline \hline
    Source & z$^{(a)}$ & N$_{H}$ (10$^{20}$ atoms cm$^{-2}$)$^{(b)}$ & R$_{\rm{500c}}$(Mpc)$^{(a)}$ & M$_{\rm{500c}}$ (10$^{14}$ M$_{\odot}$)$^{(c)}$ \\
    \hline
    HCG 62 & 0.0144 & 4.81 & 0.46 & 0.24$^{(d)}$ \\
    M49/ NGC~4472 & 0.0044 & 2.63 & 0.53 & 0.29 \\
    NGC~1550 & 0.0123 & 14.20 & 0.62 & 0.49 \\
    NGC~4325 & 0.0258 & 3.54 & 0.58 & 0.23$^{(d)}$ \\
    NGC~4636 & 0.0037 & 1.40 & 0.35 & 0.16 \\
    NGC~5044 & 0.0090 & 5.28 & 0.56 & 0.29 \\
    NGC~5813 & 0.0064 & 1.44 & 0.44 & 0.17 \\
    NGC~5846 & 0.0061 & 4.19 & 0.36 & 0.24 \\
    \hline \hline
    \end{tabular}
    \caption{Properties of the CHEERS sample targets used in this work. For more details on the relevant observations we refer to \citet{Mernier16a}. \\ \textbf{Notes.} $^{(a)}$Values from \citet{Pinto15} and references therein., $^{(b)}$Values from \citet{Kalberla05}., $^{(c)}$Values from \citet{Reiprich&Bohringer02} and references therein.,$^{(d)}$Values from \citet{Lovisari15}.}
    \label{tab:sample}
\end{table*}

\subsubsection{Background Modelling}\label{sec:background}

For the modelling of the background, we split the various contributions into two distinct categories: the Astrophysical X-ray Background (AXB), referring to sources that are astrophysical in nature and thus folded by the ARF alongside the IGrM emission, and the Non-X-ray Background (NXB), which describes particles misclassified as photons by our detector, and thus not folded by the ARF.

For the AXB, we consider the contribution from the Cosmic X-ray Background (CXB) and both the thermal emission from the Milky Way (MW) and the Local Hot Bubble (LHB). The spectral model describing the AXB is defined as
\begin{equation}
    \text{Model}=\text{phabs}\times(\text{pow}_{\text{CXB}}+\text{apec}_{\text{MW}})+\text{apec}_{\text{LHB}}
\end{equation}
assuming the same column density as before for the absorption component. We choose a fixed photon index $\Gamma=1.41$ \citep{DeLuca04} for the CXB and a Solar metallicity and redshift of 0 for the MW and LHB thermal emission. While we fix the temperature of the LHB at 0.11~keV \citep{Snowden&Kuntz11}, the temperature of the MW emission is set to the best-fit temperature between 0.15~keV and 0.6~keV \citep{McCammon02} (see \ref{sec:fit})

For the NXB, we assume:
\begin{itemize}
    \item[1.] a \texttt{bknpow} component, describing the continuum, using the photon index values for the pn and MOS detectors, adopted from \citet{Mernier15}
    \item[2.] a set of Gaussians for the fluorescent instrumental lines, adopted from \citet{Mernier15}.
    \item[3.] a \texttt{pow} component, with a best-fit spectral index $\Gamma$ between 0.1 and 1.4, for possible residual soft protons, following \citet{Snowden&Kuntz11} (see \ref{sec:fit}).
\end{itemize}

\subsubsection{Fitting Strategy}\label{sec:fit}
Lacking an off-centre pointing for constraining the background components, we choose to extract a spectrum for each observation and instrument from an annular region between 0.5' and 12'. This allows us to recover the majority of the background continuum and instrumental lines, which vary across the detector. Additionally, with this scheme, we ignore the peak of the X-ray emission associated with the galaxy group core and the potential contribution by a central AGN. However, as a result, a significant amount of thermal emission from the IGrM is still present in our spectra and needs to be taken into consideration.

Each spectrum is fitted in the 0.3-12~keV energy range with the following parameters free:
\begin{itemize}
    \item[--] the normalisation of the NXB continuum (\texttt{bknpow})
    \item[--] the centre, allowed to vary by ±0.2~keV from the expected rest-frame energy; the width, allowed to vary between 0.01 and 0.2~keV; and the normalisation of all instrumental lines (\texttt{gauss})
    \item[--] the normalisation of the CXB (\texttt{pow$_\text{CXB}$})
    \item[--] the normalisation of the LHB emission (\texttt{apec$_\text{LHB}$})
    \item[--] the temperature, allowed to vary between 0.15 and 0.6~keV, and normalisation of the MW emission  (\texttt{apec$_\text{MW}$})
    \item[--] the temperature, width and abundance of a \texttt{vlognorm} model, representing the IGrM emission, allowing them to vary by 20 per cent around the best-fit values reported in \citet{Mernier17}.
    \item[--] the spectral index, allowed to vary between 0.1 and 1.4, and normalisation of the residual soft protons (\texttt{pow}) 
\end{itemize}
Due to the large number of free parameters, we choose to fit them gradually by freeing them one by one in the listed order, ensuring a stable fit. We note that the priors on the IGrM emission have been carefully verified and proven to be a reasonable description of each system's properties by \citet{Mernier17}. The 20 per cent variation is chosen as a reasonable range to allow for slight differences between the two studies. This prior has been implemented in order to guide the fit and avoid possible local minima, considering the large number of free parameters. Repeating the background modelling using less informed priors with broader variations leads to values that remain within their respective statistical uncertainties.

To fit the spectra from the individual radial and azimuthal bins, we use the best-fit values of the background components for each observation, as previously derived, and proportion their normalisations to each bin's area. Before fitting, we ensure that the observed and modelled 10-12~keV fluxes are consistent within their statistical uncertainties, otherwise the normalisation of the NXB components is appropriately scaled. We note that for NGC~4636 and NGC~5813, a significant contribution from a central AGN has been reported. As a result, we implement an additional \texttt{pow} component, following the best-fit values reported by \citet{Mernier17}, for the bins in the central 0.5' of these two targets.

When fitting for the IGrM emission in each bin, the only free parameters are the normalisation, central temperature (kT$_{\text{central}}$), thermal distribution width ($\sigma_{\text{kT}}$) and the Fe, Mg and Si abundances of the \texttt{vlognorm} model. We assume that the IGrM emission is the same across observations and detectors and therefore, we couple the respective parameters across all spectra. 

We calculate the statistical uncertainties of the kT$_{\text{central}}$, $\sigma_{\text{kT}}$ and Fe abundance at the 1$\sigma$ level by sampling the posterior parameter distributions via the \texttt{error} command. We note that best-fits with $\sigma_{\text{kT}}$ that tend to 0 are almost always poorly constrained with high associated statistical uncertainties.

\begin{figure*}
    \centering
    \includegraphics[width=\linewidth]{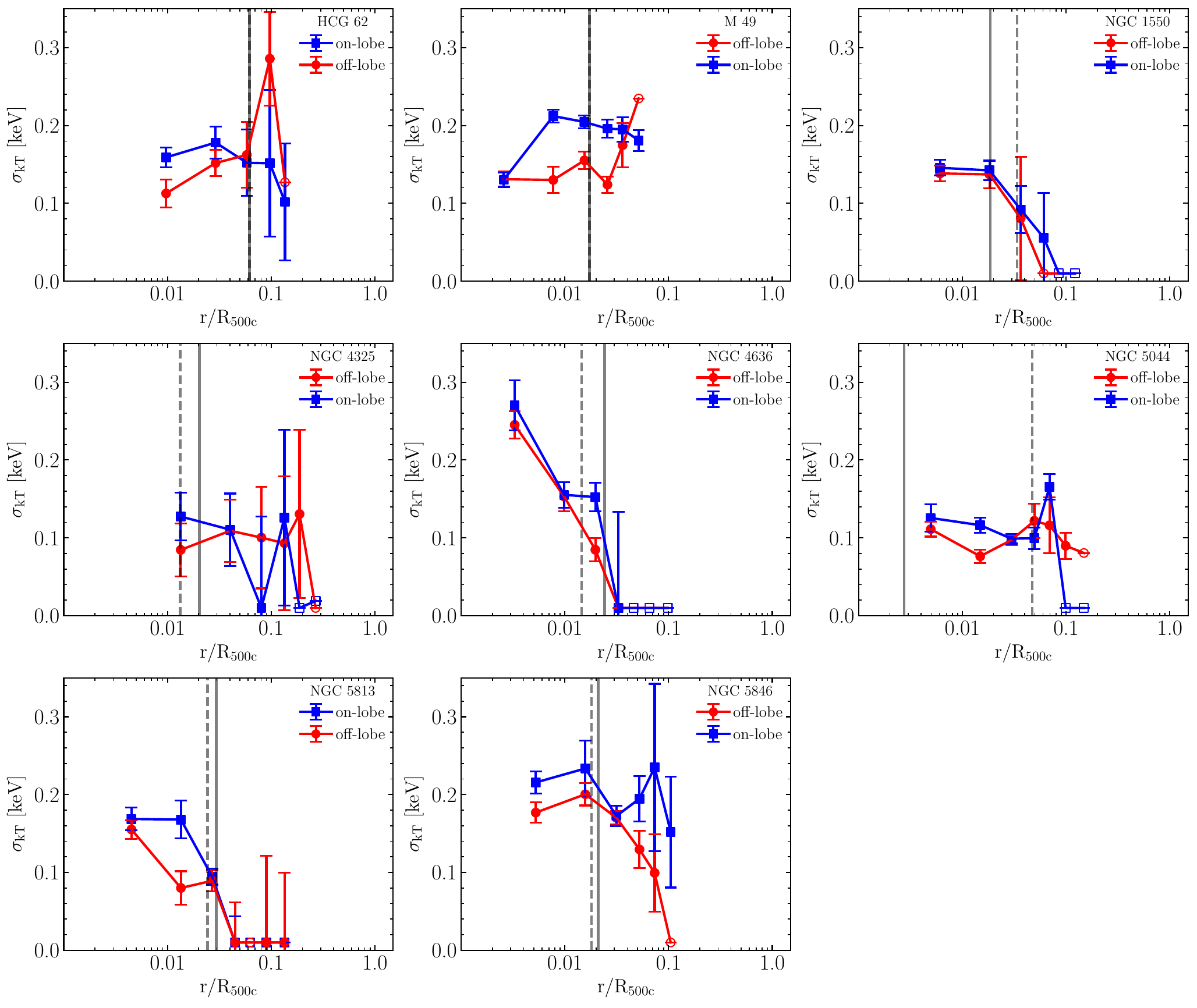}
    \caption{Average radial $\sigma_{\text{kt}}$ profiles. Data points represent the weighted averages of the thermal distribution's width in the on- and off-lobe directions, according to their relative contribution on the spectrum. Open symbols indicate error bars much larger than the axis limit. The vertical solid and dashed lines indicate the greatest extent of the radio lobes, or X-ray detected cavities if not available. Our results indicate systematically broader thermal distributions for the on-lobe directions, downstream (inside) of the radio lobes and cavities. For larger radii the width of the thermal distribution tends to vary from target to target.}
    \label{fig:sigma_weight}
\end{figure*}

\begin{figure*}
    \centering
    \includegraphics[width=\linewidth]{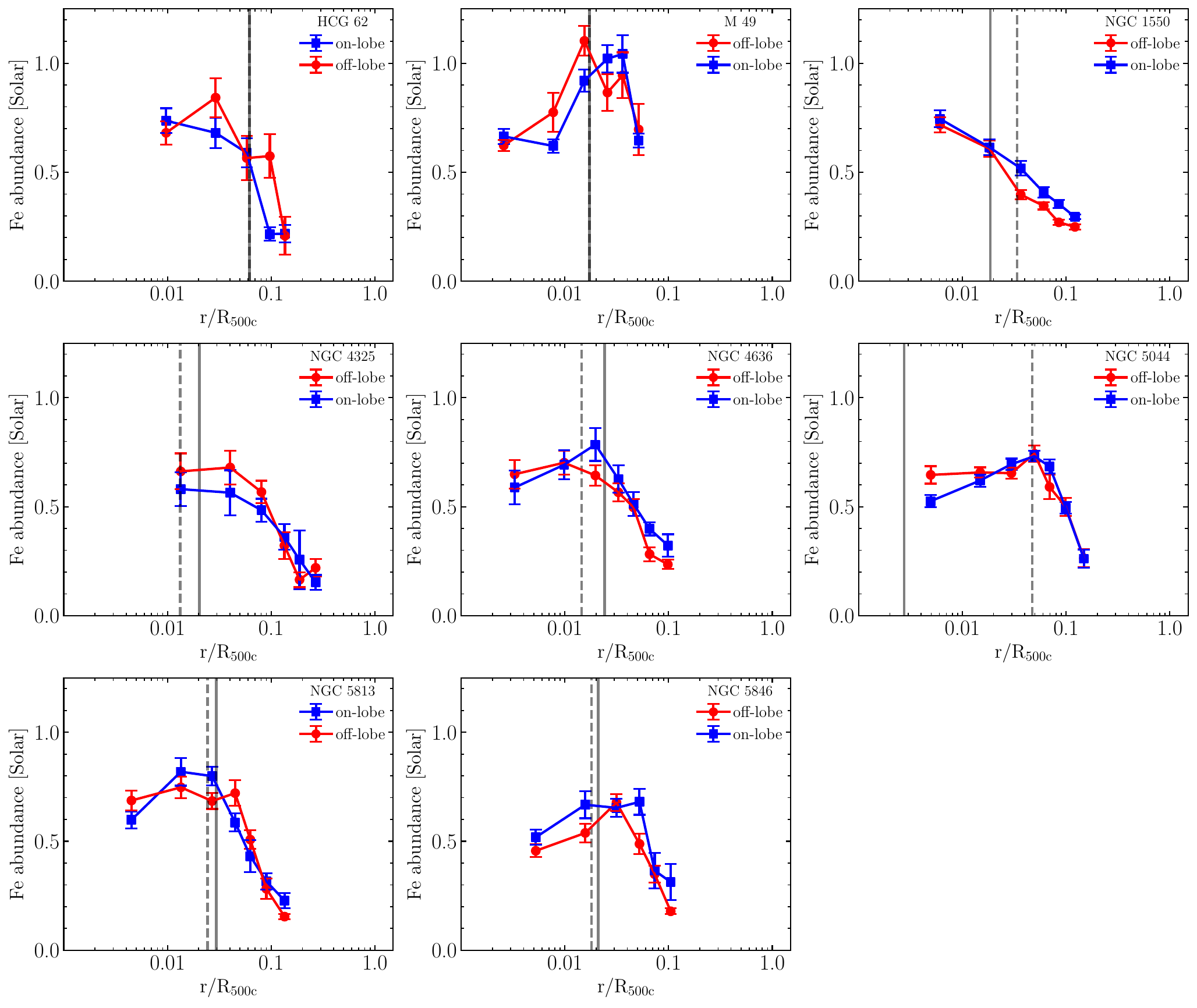}
    \caption{Average radial Fe abundance profiles. Data points represent the weighted averages of the Fe abundance in the on- and off-lobe directions, according to their relative contribution on the spectrum. The vertical solid and dashed lines indicate the greatest extent of the radio lobes, or X-ray detected cavities if not available. Our results indicate that there's little to no directional difference in the amount of Fe present in the IGrM, based on the presence or absence of AGN feedback.}
    \label{fig:Fe_weight}
\end{figure*}

\subsection{Systematic Uncertainties}
For each bin, we also estimate the systematic uncertainties on the $\sigma_{\text{kT}}$ and Fe abundance induced by the background modelling and cross-calibration between the PN-only and MOS-only measurements.

As we have already mentioned, the background parameters remain frozen during the individual bin fits and, as a result, their uncertainties are not accounted for in the statistical errors. We choose to test the robustness of the best-fit values by varying the free background parameters in Sect. \ref{sec:fit}. Namely, we perform a spectral fit on each bin for a total of 30 times after randomly sampling the background values for each parameter from a Gaussian distribution, described by their respective best-fit value and 1$\sigma$ error. We use the standard deviations of the best-fit $\sigma_{\text{kT}}$ and Fe abundance samples as our systematic uncertainties due to the the background modelling. We note that the systematics increase in the low surface brightness regions but remain smaller than the statistical errors.

Another possible source of systematic errors are the uncertainties due to cross-calibration of the PN and MOS spectra. In order to identify possible tensions between the two instruments, we fit each bin's MOS and pn observations separately. We assume that the offset between the MOS-only and pn-only best-fit values is representative of the systematics due to cross-calibration. We note that, similar to the systematics due to the background modelling, the uncertainties are not statistically significant in the brighter regions but, they become the primary source of errors in the low surface brightness regions.

In order to reflect the statistical significance of the systematics in our analysis, we choose to always adopt the largest source of uncertainty as the associated error for each bin. Namely, we increase the error bars for each $\sigma_\text{kT}$ and Fe abundance measurement until they cover the extent of their respective systematic errors, if necessary. We also attempt to preserve any possible asymmetries in the posterior distribution by adjusting the upper and lower errors independently.

\section{Results}\label{sec:results}
Starting with the thermal structure of the IGrM, in \autoref{fig:sigma_weight} we present the radial $\sigma_\text{kT}$ profiles for all directions of our targets. We choose to weight the reported thermal distribution widths for each bin by a factor of 1/$\sigma_{\rm{\sigma_{kT},i}}^2$, where $\sigma_{\rm{\sigma_{kT},i}}$ the error on each bin's temperature width, before averaging for the two directions, following \citet{Mernier17}. With this scheme, we effectively weight each bin with respect to its emission measure, which accounts for each bin's relative contribution to the average due to size and brightness. We present the radial profiles for all individual directions in \autoref{fig:sigma_indiv}.

As \autoref{fig:sigma_weight} suggests, the inner regions, where we assume that the current AGN feedback is interacting with the IGrM, have systematically broader thermal distributions in the on-lobe directions then their off-lobe counterparts. The offset between the two varies strongly from target to target, ranging from 1 to 4$\sigma$ in the radial bins before the largest radio lobe extent. However, it appears to be independent of the gas temperature as we find little to no difference between the two directions (see \autoref{fig:kT_weight})

We also note that, for the majority of our targets, the width of the thermal distribution tends to 0, or becomes unconstrained, towards the edge of the core. While we observe this trend for both the on- and off-lobe directions, the later are more likely to have a statistically significant amount of multi-phase gas, reflected in the $\sigma_{\text{kT}}$, at the edge of the core.

Despite the clear differences in the width of the thermal distribution between the two directions, a similar asymmetry is not evident in the Fe abundance profiles. In \autoref{fig:Fe_weight}, we present the on- and off-lobe Fe radial profiles, weighted with respect to each bin's Fe abundance error, following a similar scheme as before. As our profiles indicate, there is little to no systematic differences in the the Fe abundance of the two directions. Specifically for the inner regions, where we observe the larger offsets in the $\sigma_{\text{kT}}$ between the on- and off-lobe directions, the Fe abundance differs by less than 2$\sigma$ throughout our sample, with the exception of two bins in HCG~62 and NGC~1550 where a 3.4 and 3.9 $\sigma$ tension is found, respectively. Overall, this seems to suggest a lack of any strong directional abundance enhancement, despite the evidence of ongoing AGN feedback.

The overall shape of the Fe abundance profiles is consistent with the ones presented in \citet{Mernier17} and have a large target-to-target variation. While M49, NGC~4636 and NGC~5813 show evidence of a central enhancement peak in both on- and off-lobe directions, the rest of our targets have relatively flat Fe abundance profiles, lacking any such evidence. A notable exception is NGC~1550 where we observe a sharply deceasing Fe abundance across the core.

\section{Discussion}\label{sec:discussion}
\subsection{Effects of AGN feedback on the IGrM}
\begin{figure}
    \centering
    \includegraphics[width=\linewidth]{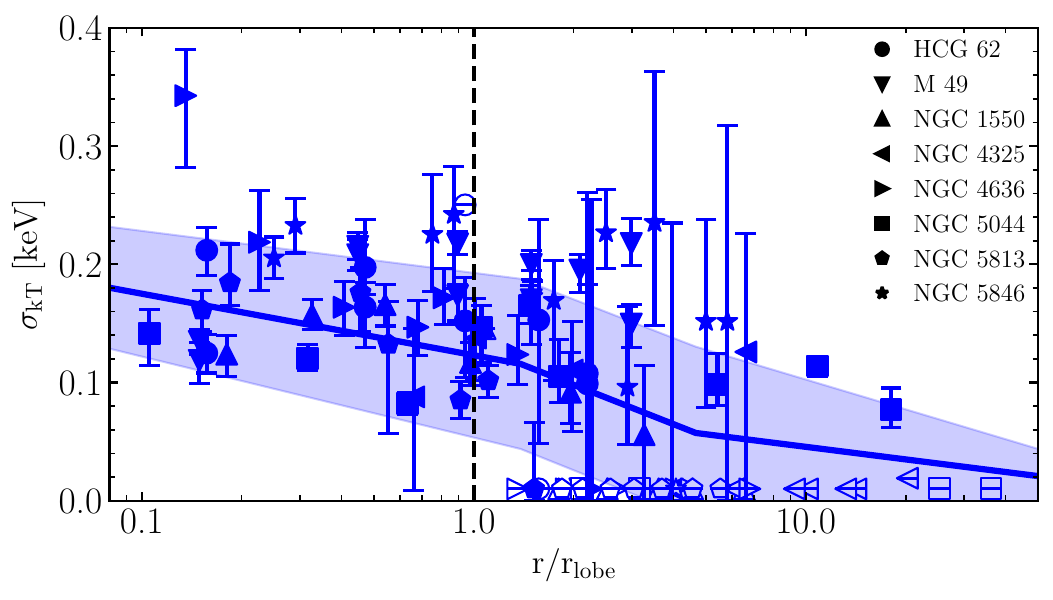}
    \caption{The width of the thermal distribution in the individual on-lobe directions, based on their relative position to the largest extent of the radio lobes or X-ray detected cavities. Open symbols indicate error bars much larger than the axis limit. The vertical dashed line represents the largest radial separation of the manifestations of AGN feedback. The solid blue line indicates the weighted running average of our measurements and the shaded area the 1$\sigma$ scatter. Our results suggest the presence of  a non-negligible amount of multi-phase gas upstream from the current manifestations of AGN feedback.}
    \label{fig:sigma_lobe}
\end{figure}

\begin{figure}
    \centering
    \includegraphics[width=\linewidth]{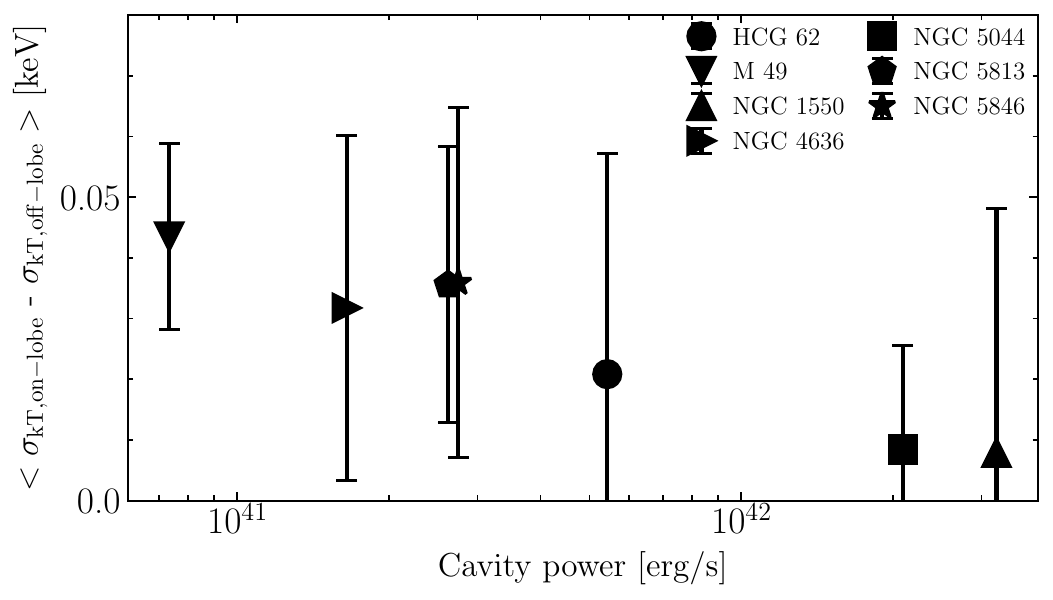}
    \caption{The average thermal width excess on the on-lobe directions, with respect to the off-lobe, compared to the average cavity power of each target. Data points represent the average difference in the widths of the thermal distribution downstream across our sample. Our results indicate a possible anti-correlation with cavity power, suggesting a more isotropic distribution of cooler gas in systems with more powerful AGN feedback.}
    \label{fig:sigma_cav}
\end{figure}

\subsubsection{Thermal structure}
Our results indicate that AGN feedback has a clear effect on the thermal structure of the IGrM. To our knowledge, this is the first time that a systematic asymmetry is reported in the width of the thermal distribution, with more multi-phase gas found along the direction of the radio lobes. 

In order to quantify its effect, we choose to further examine the radial $\sigma_\text{kT}$ profiles of the on-lobe directions with respect to the largest extent of their corresponding radio lobes. This allows us to highlight any differences in the width of the thermal distribution between regions that are and are not currently impacted by AGN feedback, as indicated by the presence of radio lobes and X-ray cavities, along the same direction. In \autoref{fig:sigma_lobe} we see that, on average, downstream (inner) regions have broader thermal distributions than their upstream (outer) counterparts. We also find a steeper slope on the running average when transitioning from downstream to upstream, that similarly indicates a connection between multi-phase gas and the radio lobes.

However, it is evident that there are many regions upstream with a significant amount of multi-phase gas. Even for cases approximating a single temperature distribution, $\sigma_\text{kT}$ remains for the most part unconstrained, which means that we can not rule out the presence of cooler gas. While these areas should be unaffected by the current AGN feedback, as indicated by the largest extent of the observed radio lobes, this does not mean that they could not have been affected by previous AGN outbursts. It is possible that such events could have uplifted central gas, in a fashion similar to the current AGN feedback, which can still be seen in the $\sigma_\text{kT}$ profiles. If that is the case, careful examination of the multi-temperature structure of the IGrM could be key in understanding the history of the central AGN in these systems. However, such an analysis exceeds the scope of this work.

Another interesting feature of the averaged $\sigma_{kT}$ profiles is the offset in the thermal distribution widths between the on- and off-lobe directions. While, as we have already pointed out, the on-lobe directions have systematically broader thermal distributions than their off-lobe counterparts, the difference between the two varies from target to target.

In order to quantify this effect, we choose to examine the average $\sigma_{\text{kT}}$ excess, calculated as the average difference in the $\sigma_{\text{kT}}$ between on- and off-lobe directions in all downstream radial bins, for our entire sample. Assuming that a broader thermal distribution implies that large amounts of cooler gas are present, the average $\sigma_{\text{kT}}$ excess highlights its distribution, with a value of 0 implying isotropy. 

In \autoref{fig:sigma_cav} we present the average $\sigma_{\text{kT}}$ excess as a function of each target's average cavity power ($\rm{P}_{cav}$), which we use as a proxy of the AGN power. The latter is estimated following \citet{Panagoulia14}, using their reported cavity properties and pressure profiles from the literature, when available. Our results indicate an anti-correlation between the two, with a Spearman correlation coefficient $r_S=-0.86$ (p-value $\sim$ 0.014), implying that the distribution of cooler gas is more isotropic in systems with stronger AGN feedback. However, both the excess $\sigma_{\text{kT}}$ and the cavity power, are associated with large uncertainties.

We speculate that this trend could be indicative of a more efficient mixing, induced by the AGN feedback. The shallower gravitational potential wells of galaxy groups could suggest that the energy injections from the central SMBH are capable of not only uplift of cooler central gas, but also diffusion in the azimuthal axis. In such a scenario, larger energy injections can lead to a smaller $\sigma_{\text{kT}}$ excess. However, due to the large uncertainties of our measurements, the observed behaviour can be subjected to target-to-target variation and not representative of a broader global trend.

However, it is important to note that while the trend appears to be both strong and statistically significant at the $\alpha$=0.05 level of significance, both the excess $\sigma_{\text{kT}}$ and the cavity power, are associated with large uncertainties. Therefore, it is possible that the observed behaviour is subjected to target-to-target variation and not representative of a broader global trend.

Overall, while there is a clear asymmetry in the thermal structure of the gas, due to AGN feedback, our results imply a more complex dynamical state of our targets' IGrM, despite their overall relaxed morphology. This highlights the need for a more detailed study of those low mass systems, both theoretically and observationally.

\begin{figure}
    \centering
    \includegraphics[width=\linewidth]{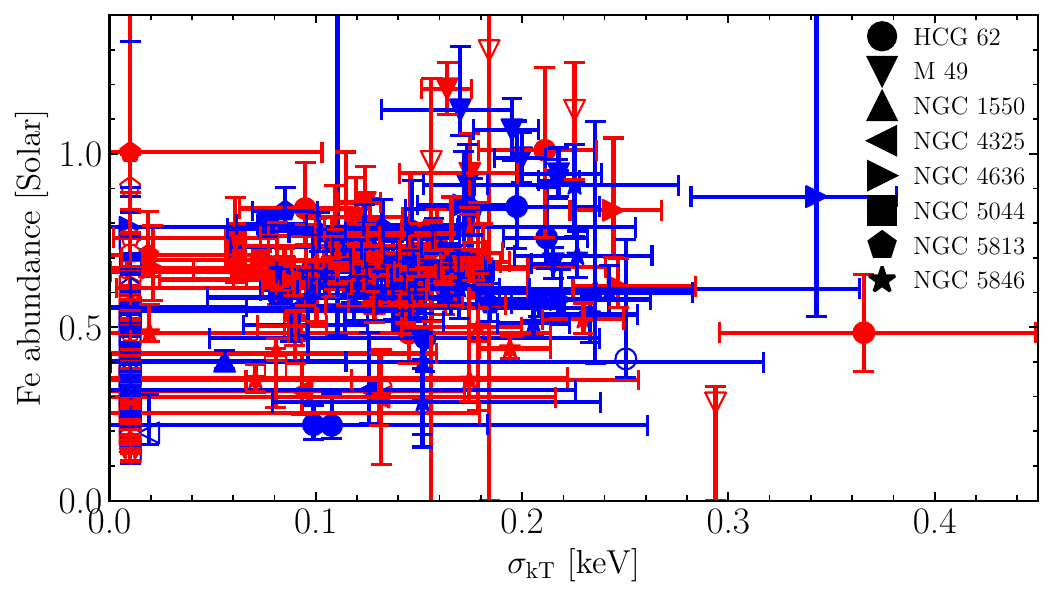}
    \caption{The Fe abundance in each individual bin vs. its thermal distribution's width. Bins are colour coded based on their location with respect to the direction of the AGN feedback (on-lobe in blue, off-lobe in red). Open symbols indicate error bars much larger than the axis limit. Our results highlight the decoupling between the width of the thermal distribution, shown to be tracing the AGN feedback, and Fe abundance with no clear correlations between the two.}
    \label{fig:Fe_sigma}
\end{figure}

\begin{figure*}
    \centering
    \includegraphics[width=\linewidth]{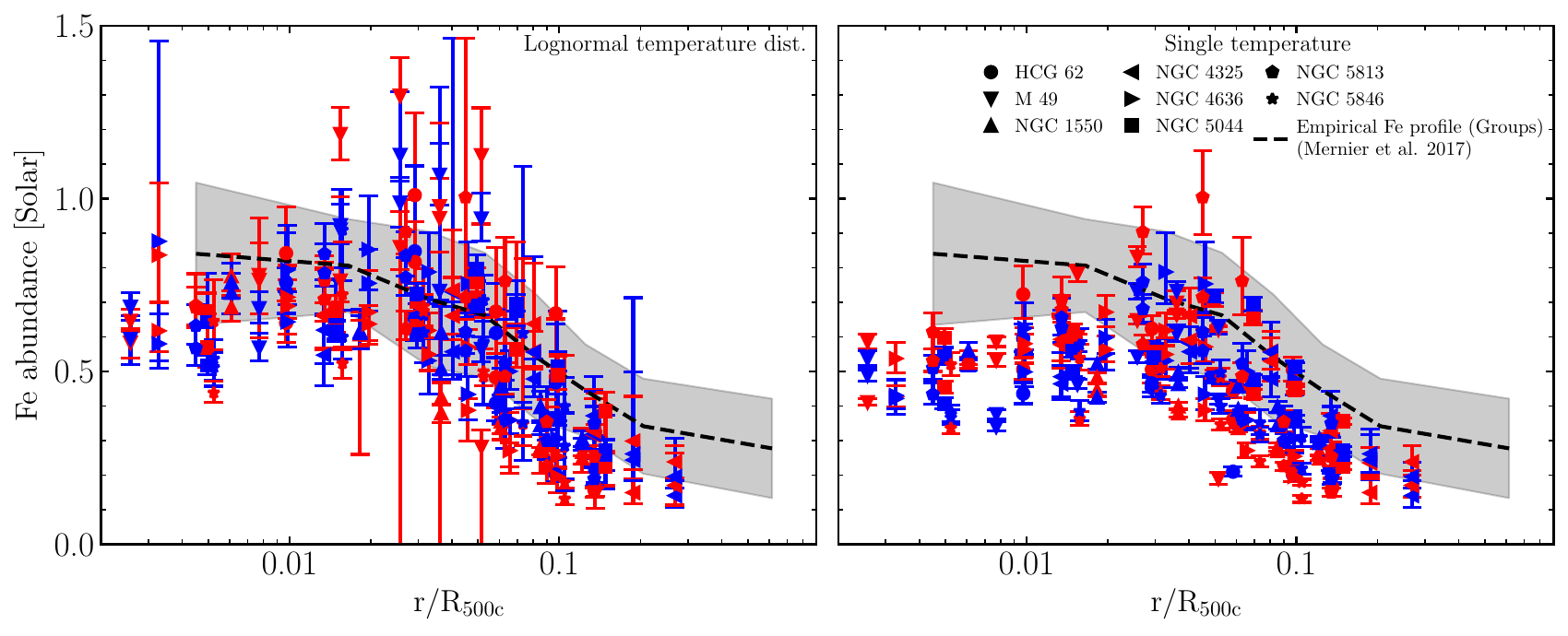}
    \caption{Fe abundance radial profiles for all individual directions in our sample. We compare the results of this work (left) with the Fe abundances assuming only a single temperature model throughout (right). Blue data points illustrate profiles in the on-lobe directions, while the off-lobe ones are represented with red. The empirical Fe abundance profile for galaxy groups from \citet{Mernier17} is highlighted by the black dashed line. While ignoring the multi-phase gas, leads to systematically lower Fe abundances in the centres of our targets, we find no strong systematic differences in the on- and off-lobe directions.}
    \label{fig:Fe_all}
\end{figure*}

\subsubsection{Chemical structure}
While the AGN feedback appears to have a directional effect on the thermal structure of the IGrM, this is not the case for the chemical structure. As we have previously shown, we find little to no differences in the Fe abundance radial profiles of the two directions, regardless of the thermal structure. In fact, as \autoref{fig:Fe_sigma} suggests, the two quantities are likely decoupled, outside of a possible mutual anti-correlation with radius.

However, the isotropic Fe abundance distribution of the IGrM across our sample, contradicts the diversity of directional AGN feedback effects on its chemical structure found in literature. For example, previous studies have reported depressions in the Fe abundance along the directions of uplifted gas dragged by AGN feedback \citep[e.g.][]{GendronM17,Su19}, that we do not find, even though these systems are included in our sample. One likely explanation for such differences, as it was demonstrated in \citet{Randall15} for NGC~5813, could reside in the Fe bias \citep[][]{Buote99}, where one can systematically underestimate the true Fe abundance if a multi-phase medium, such as the IGrM, is approximated by a single temperature model.

In \autoref{fig:Fe_all} we showcase this effect by calculating the Fe abundance profiles of our targets, assuming a single temperature model throughout. We find that, as expected, the central Fe abundances are systematically below the empirical Fe profile from \citet{Mernier17}, as they represent regions with large amounts of multi-phase gas. While HCG~62 and NGC~5813 (shown in \autoref{fig:Fe_1T}), exhibit an abundance enhancement in the off-lobe directions, the effect is not systematic throughout our sample. Almost all remaining targets in our sample show little to no difference in the amount of Fe between directions. This suggests that additional effects must be in play in order to explain the previously reported diversity of chemical IGrM structures.

Cases were a strong Fe abundance enhancement has been found to be associated with the AGN feedback \citep[e.g.][]{O'Sullivan05,O'Sullivan11,Lagana15}, can not be a product of the Fe bias. However, given the consistency of the Fe abundance radial trends found in this work, we speculate that this could be an artifact due to atomic library incompleteness. Small changes in the atomic codes and plasma emission models can have significant differences in the shape of the X-ray emission, especially in the Fe-L complex. As a result, it is possible that a model can overestimate the amount of Fe present, in order to compensate for a lower predicted emissivity. 

Since the effects of the Fe bias are a direct product of limited spectral resolution and the biases due to atomic libraries can be highlighted when comparing with high resolution spectra, the previously mentioned tensions can be addressed by resolving the Fe-L bump. Overall, our results (re-) open the question on the diversity of chemical structures in the IGrM in relation with central AGN feedback, with a consistent description of the Fe abundance distribution as isotropic. Such a description appears to be more in favour of an early, instead of a late, enrichment scenario, with the IGrM being formed from gas that is already enriched and well-mixed. However, a deeper understanding of the interaction between the central AGN and the hot atmospheres of these systems is necessary from both observations and theoretical works. Additionally, the distribution of Fe appears to be insensitive to the directionality of the radio lobes, questioning both its reliability to trace AGN feedback in galaxy groups.

\subsection{Comparison with clusters}
As stated previously, studies of the ICM have indicated that clusters systematically show an increase in the amount of Fe along the direction of radio lobes and cavities \citep[e.g.][]{Kirkpatrick09,Kirkpatrick11}. This however is in tension with what our findings regarding the IGrM.

One possibility is that the Fe enhancements in clusters are artificial. As we have previously mentioned, improper modelling of the plasma emission or of the thermal distribution of the underlying gas \citep["inverse Fe-bias"; e.g][]{Rasia08,Simionescu09,Gastaldello10} could overestimate the amount of Fe present in each direction. Similarly, it is possible that the lack of an abundance enhancement is subjected to systematics due to the reliance on the Fe-L complex. Unlike galaxy clusters where the continuum is purely thermal and the ICM is hot enough to excite the Fe K$\alpha$ line, spectral analysis of the IGrM relies on emission from the Fe-L complex, which is both unresolved at CCD spectral resolution and subjected to degeneracies with a continuum that is no longer purely thermal. Therefore, it is possible for any anisotropies in the chemical structure of the IGrM to be erased due to such systematics.

Either scenario would suggest that both the ICM and the IGrM show similar levels of metal mixing. However, \citet{Riva22} have indicated only a modest level of systematic uncertainties in the Fe abundance of intermediate mass galaxy clusters due to their reliance on the Fe-L complex. Assuming that such errors remain comparable for lower mass systems, the latter scenario seems unlikely. Overall, higher resolution spectroscopy and a more sophisticated modelling of the 2-4~keV plasma could resolve this tension and provide us with an accurate representation of the underlying gas' chemical properties.

Alternatively, it is possible that an Fe abundance enhancement in our galaxy groups is present along the on-lobe directions, but its detection remains limited due to spatial resolution, modelling assumptions and photon statistics. For example, throughout our analysis we assume that all thermal phases in our model have the same Fe abundance. While this assumption is consistent with previous works in clusters, it is possible that it might not hold for galaxy groups. As a result, we might be missing a cooler, metal rich gas component in the on-lobe directions. Similarly, our model uses a single point estimate for the Fe abundance which is sensitive to the mean Fe abundance along our line of sight. Considering the sizes of our bins, it is likely that a thin and less bright stream of overabundant gas gets averaged out by the bulk of the IGrM, resulting in its non-detection. Higher spatial-spectral resolution and sensitivity, as well as studies of the IGrM from theoretical works, can significantly improve our capability to both resolve and predict the actual amount of Fe present inside those regions.

Finally, it is likely that the tension between the ICM and IGrM with respect to their chemical structure is a product of their halo masses. Even though the energy injected to both the ICM and the IGrM is comparable, due to the similarly massive SMBHs, galaxy groups and giant ellipticals reside in significantly less massive halos. As a result, AGN feedback can remove a large fraction of baryons and the metals associated with them from the halos of galaxy groups. This would negate any possible enhancement via uplifted metals, while mixing the IGrM efficiently early on. This notion appears to be supported by theoretical works, that highlight the eradication of any abundance asymmetries in halos of log$_{10}$(M$_{200c}$/M$_\odot$)~13.0 \citep[e.g.][]{Truong21}. This scenario could explain the flatness of the abundance peak in the profiles of galaxy groups and the decoupling of the thermal distribution width and Fe abundance, as the former is associated with the current AGN activity while the latter has already been mixed. However, due to the limitations we mentioned previously a greater understanding of the effects of AGN feedback in these lower mass systems is necessary both from observations and simulations.

\section{Conclusions}\label{sec:conclusions}
In this paper, we have re-examined archival \textit{XMM-Newton} data for a subset of galaxy groups and giant elliptical galaxies from the CHEERS sample. The scope of this work is to investigate the effects of AGN feedback on the surrounding hot atmosphere, namely on its thermal structure and metal distribution, of such lower mass systems in systematic way. The main difference between our analysis and previous published works using a similar dataset is that we examine those effects by factoring in the directional nature of AGN feedback while implementing a more realistic description of the thermal structure via a Log-Normal distribution. Our main results are summarised as follows.

\begin{itemize}
    \item[-] We systematically find an extended multi-temperature structure along the direction of the observed radio-lobes, or the position of known cavities, across our entire sample. Evidence of multi-phase in the off-lobe directions is mainly found closer to the cores and can be a product of gas mixing on those scales.
    
    \item[-] Our results suggest that recent AGN feedback increases the amount of multi-phase gas in the IGrM, as indicated by broader thermal distribution widths in the on-lobe directions. However, we find a non-negligible amount of cooler gas towards the outskirts, most likely due to former energy injections.

    \item[-] Despite the anisotropic distribution of the multi-phase gas, we find no directional differences in the Fe abundance radial profiles. Rather, they indicate a similar distribution of Fe across azimuths, consistent with the empirical Fe profile from \citet{Mernier17}.
\end{itemize}
Overall, our analysis indicates that an isotropic distribution of metals in the cores of galaxy groups and ellipticals is a common occurrence. This is despite the asymmetries in the distribution of the multi-phase gas suggesting a possible efficient mixing of metals in the IGrM. However, our understanding of the dynamical state and chemical enrichment of the IGrM still remains limited due to spectral and spatial resolution. Future missions with higher spectral resolution and sensitivity, combined with dedicated theoretical works in this mass scales would allow us to better understand the role AGN feedback plays in the distribution of metals in the IGrM.

\section*{Acknowledgements}
We thank the reviewer for their constructive and useful suggestions. We thank Simona Giacintucci and Konstantinos Kolokythas for sharing the GMRT images for our targets. DC is a fellow of the International Max Planck Research School for Astronomy and Cosmic Physics at the University of Heidelberg (IMPRS-HD) and acknowledges funding from the European Union (ERC, COSMIC-KEY, 101087822, PI: Pillepich). The scientific results reported in this article are based in part on data obtained from the XMM-Newton Science Archive (XSA). SRON Netherlands Institute for Space Research is supported financially by the Netherlands Organisation for Scientific Research (NWO). The material is based upon work supported by NASA under award number 80GSFC21M0002.

%%%%%%%%%%%%%%%%%%%%%%%%%%%%%%%%%%%%%%%%%%%%%%%%%%
\section*{Data Availability}

The data underlying this article were accessed from the XMM-Newton Science Archive (XSA). The GMRT data used in this article were provided by Simona Giacintucci and  Konstantinos Kolokythas by permission. Data will be shared on request to the corresponding author with their permission. The derived data generated in this research will be shared on reasonable request to the corresponding author.

\bibliographystyle{mnras}
\bibliography{ms}

%%%%%%%%%%%%%%%%%%%%%%%%%%%%%%%%%%%%%%%%%%%%%%%%%%

%%%%%%%%%%%%%%%%% APPENDICES %%%%%%%%%%%%%%%%%%%%%

\appendix

\section{Individual profiles}
In this section we present the radial $\sigma_{\text{kT}}$ and Fe abundance profiles for all individual directions, across our sample. Since the extent of the radio lobes or the location of the X-ray detected cavities is not symmetric, we annotate them with the line style of the on-lobe direction they belong to.
\begin{figure*}
    \centering
    \includegraphics[width=\linewidth]{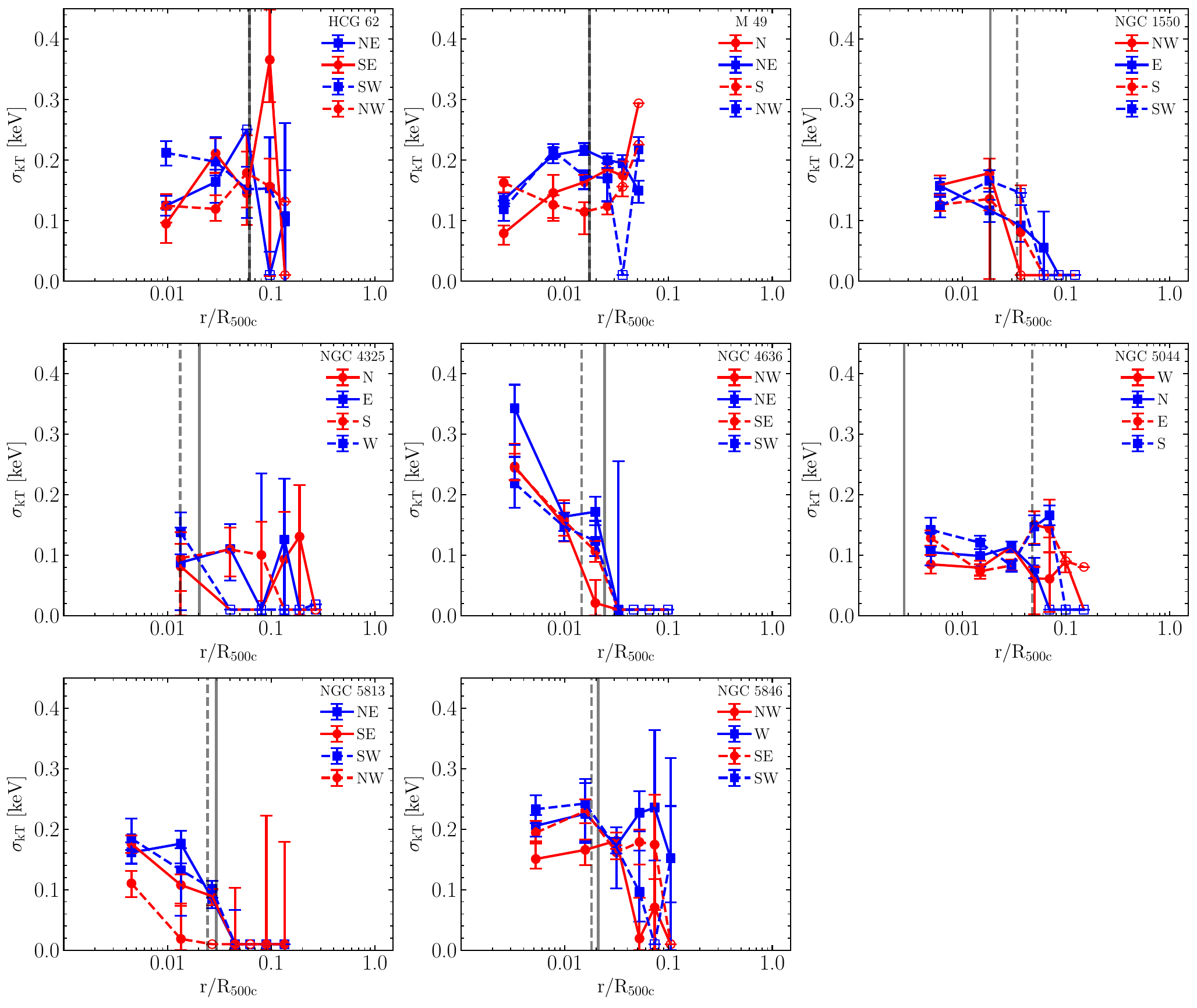}
    \caption{Radial profiles of the thermal distribution's width for all directions. Open symbols indicate error bars much larger than the axis limit. The vertical solid and dashed lines indicate the greatest extent of the radio lobes, or X-ray detected cavities if not available, for the solid and dashed on-lobe directions respectively.}
    \label{fig:sigma_indiv}
\end{figure*}

\begin{figure*}
    \centering
    \includegraphics[width=\linewidth]{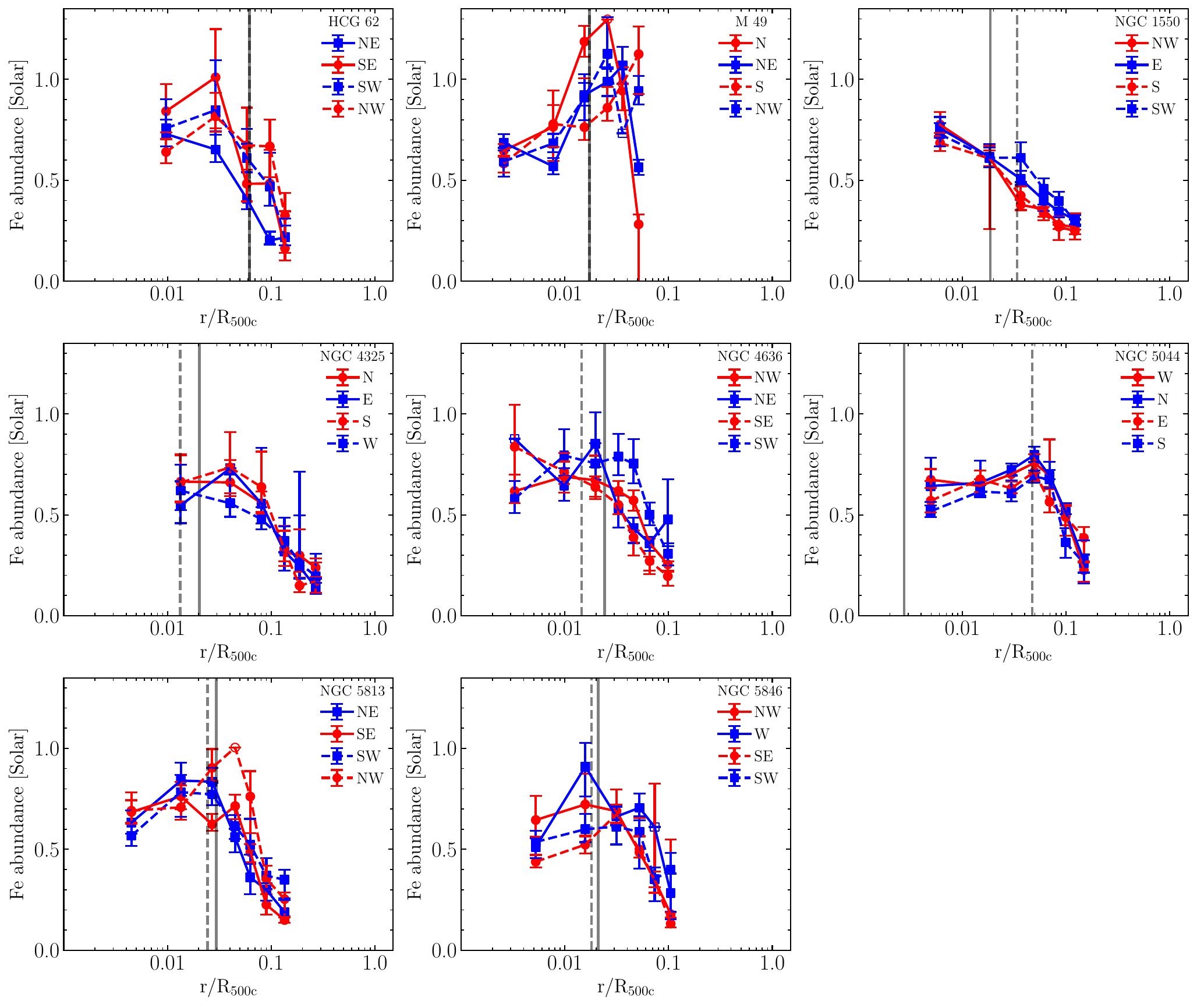}
    \caption{Radial profiles of the Fe abundance for all directions. The vertical solid and dashed lines indicate the greatest extent of the radio lobes, or X-ray detected cavities if not available, for the solid and dashed on-lobe directions respectively.}
    \label{fig:Fe_indiv}
\end{figure*}

\section{Model systematics}\label{sec:systematics}
In this section we examine possible effects that can be biasing the interpretation of our trends. A possible bias, inflating the difference in the width of the thermal distribution between on- and off-lobe directions can be the \texttt{vlognorm} model's description of the thermal structure. If the peak temperature of the gas deviates strongly between the two directions, it is possible for the model to inflate the width of the distribution to compensate for it.

\begin{figure*}
    \centering
    \includegraphics[width=\linewidth]{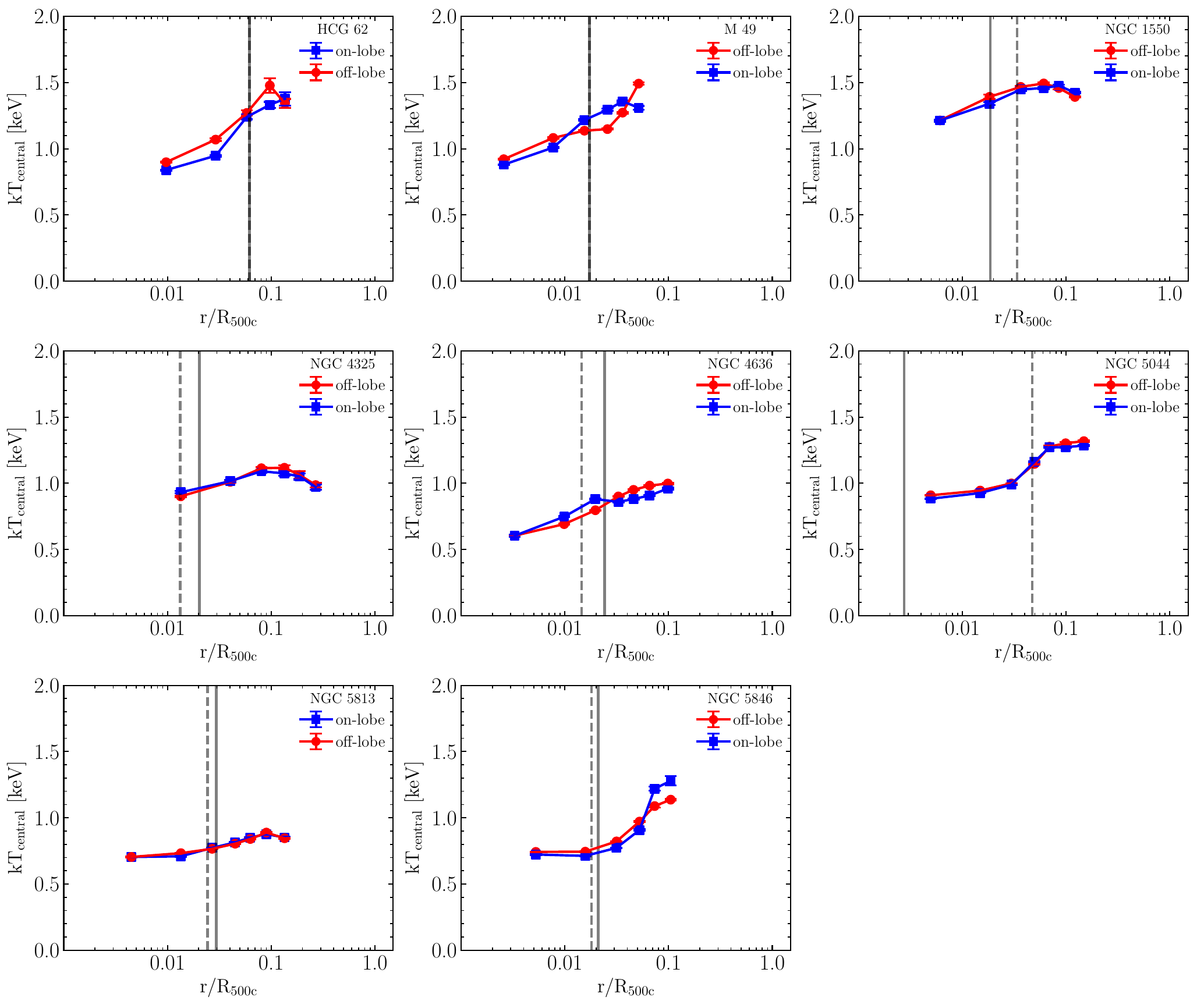}
    \caption{Weighted average radial profiles of the thermal distribution's centre for the on- and off-lobe directions. The vertical solid and dashed lines indicate the greatest extent of the radio lobes, or X-ray detected cavities if not available. There are effectively no differenced in the central temperature between the two directions suggesting that it does not play a role in the different thermal widths.}
    \label{fig:kT_weight}
\end{figure*}

In \autoref{fig:kT_weight} we present the weighted average temperature between the on- and off-lobe directions. It is evident that, the temperature gradient remains consistent between them, suggesting that the broader widths reported in the on-lobe directions are physical.

\begin{figure*}
    \centering
    \includegraphics[width=\linewidth]{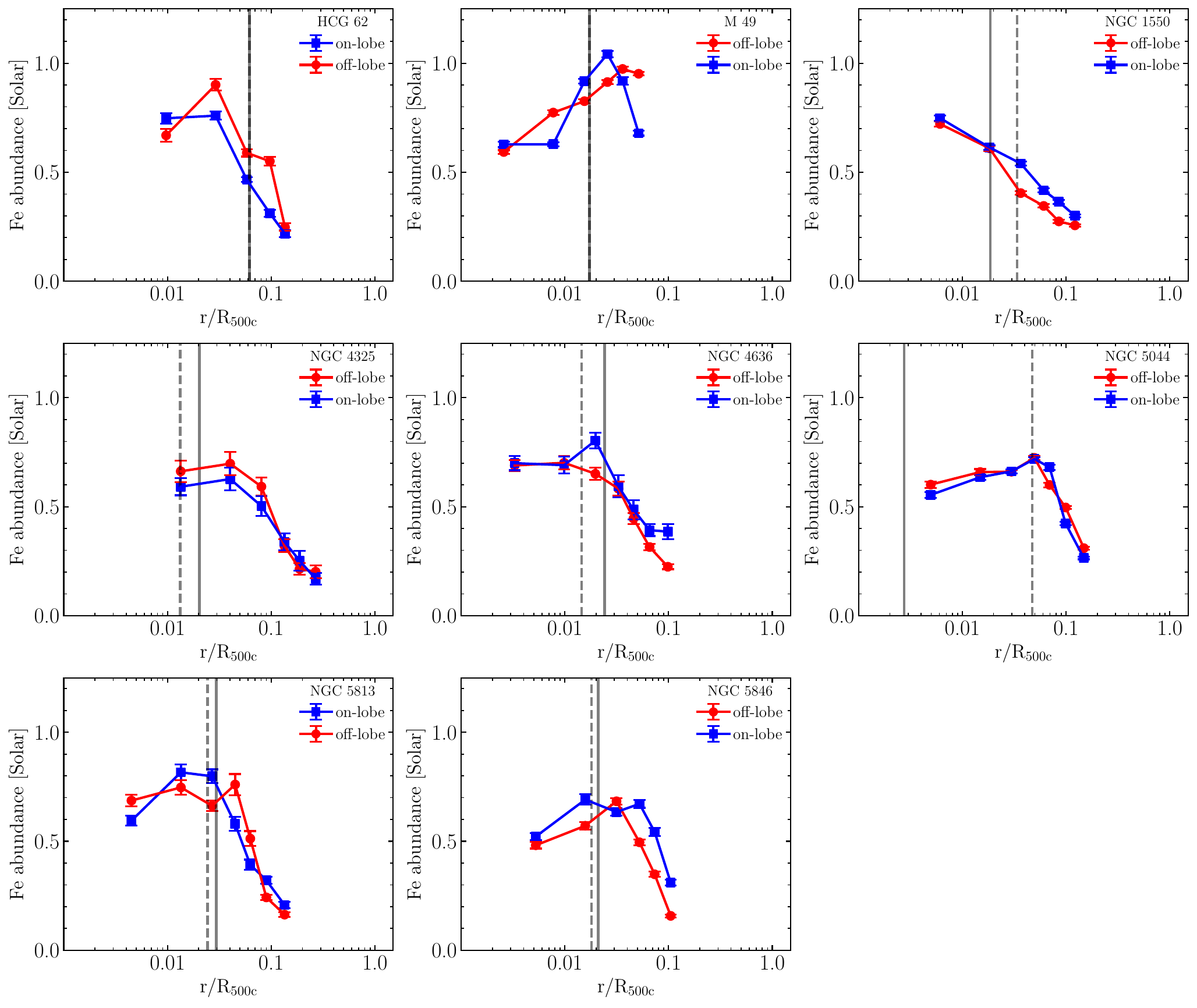}
    \caption{Weighted averaged Fe abundance profiles for the on- and off-lobe directions, assuming a single temperature model throughout. The vertical solid and dashed lines indicate the greatest extent of the radio lobes, or X-ray detected cavities if not available. Our results suggest that ignoring the multi-phase nature of the IGrM can lead to cases where the Fe abundance of the on-lobe directions is depressed to the extent of an off-lobe enhancement. However, the rest maintain an azimuthally symmetric chemical structure despite the reduced inferred Fe abundance.}
    \label{fig:Fe_1T}
\end{figure*}

Finally, we want to quantify the extend to which the Fe bias could influence our inferred Fe abundance profiles. In \autoref{fig:Fe_1T} we present the weighted average Fe abundance profiles for the two directions, modelling the gas with a single temperature thermal model. It is evident that the galaxy groups and giant ellipticals with the highest amount of multi-phase gas in their centres are heavily subjected to the Fe bias, resulting in enhancements in the off-lobe directions. However, the rest of our targets maintain the view of a symmetric chemical structure across azimuths, even if appearing metal poorer. 

%%%%%%%%%%%%%%%%%%%%%%%%%%%%%%%%%%%%%%%%%%%%%%%%%%

% Don't change these lines
\bsp	% typesetting comment
\label{lastpage}
\end{document}